\documentstyle[a4wide,12pt,epsfig]{article}
\title{Computing the noncomputable}
\author{Tien D Kieu\footnote{email: kieu@swin.edu.au},\\
Centre for Atom Optics and Ultrafast Spectroscopy,\\Swinburne
University of Technology, Hawthorn 3122, Australia}
\begin{document}
\maketitle
\begin{abstract}
We explore in the framework of Quantum
Computation the notion of computability, which holds a central position in
Mathematics and Theoretical Computer Science.  A quantum algorithm 
that exploits the quantum adiabatic processes is considered for the
Hilbert's tenth problem,
which is equivalent to the Turing halting problem and known to be mathematically
noncomputable.  Generalised quantum
algorithms are also considered for some other mathematical
noncomputables in the same and of different noncomputability classes.
The key element of all these algorithms is the measurability of both the
values of physical observables and of the quantum-mechanical 
probability distributions for these values.
It is argued that computability, and thus the limits of Mathematics,
ought to be determined not solely by Mathematics itself but also by physical
principles.
\end{abstract}
% {\bf Notes:  QM as one of the most successful theories / Godel's misquoted as
% the end of computability [in Arithmetic] / ...}
\newpage
\tableofcontents
\newpage
\begin{flushright}
{\em Wir m\"ussen wissen\\
Wir werden wissen}\\
David Hilbert
\end{flushright}
\section{Introduction}
This article is a brief introduction to an effort to compute the
otherwise mathematical noncomputable.  The new ingredients here 
for this effort
are those supplied by physical principles, the quantum principles in particular,
and situated outside the domain 
of Mathematics where the limits of computability are set.

In the next Section we will summarise the important concept of Turing
machines and their relationship to the mathematical recursive functions.
They will set the scene for the notion of computability as delimited by
the Church-Turing hypothesis.  We then review the noncomputable results
for the Turing halting problem and the equivalent Hilbert's tenth
problem, which is directly accessible for some quantum-mechanical exploration
in a later Section.  We also introduce some other noncomputable problems
and the Chaitin's $\Omega$ numbers before moving on to a discussion of
Quantum Mechanics and Quantum Computation in Secs.~\ref{sec:qm}
and~\ref{sec:qc}.

In Sec.~\ref{sec:qm}, we emphasise the concept of coherent states and
the quantum adiabatic theorem in order to introduce the model of Quantum
Computation by adiabatic processes in Sec.~\ref{sec:qc}.  With this
computation model we then propose a quantum algorithm for Hilbert's
tenth problem in Sec.~\ref{sec:10} (where the illustrating examples are done
in collaboration with Andrew Rawlinson), and discuss some of its finer 
points in the following Sec.~\ref{sec:dis}.  To explore the limits of 
this quantum algorithm we also consider some
generalisation for other noncomputables in Sec.~\ref{sec:qcothers}.

Before concluding with some remarks in the final Section, we present a
reformulation of Hilbert's tenth problem, as inspired by our quantum
algorithm, in terms of an infinitely coupled set of differential
equations in Sec.~\ref{sec:reform}.

\section{Effective Computability and Noncomputability}
%Computability 
%Turing machine
%Turing Universal machine
%Godel's encoding
\subsection{Turing machines}
The concept of computation is at the heart of Mathematics and hard
sciences but it was only made precise by Alan Turing relatively
recently, at the beginning of the twentieth century~\cite{turingbiblio}.  
With the introduction of theoretical and mathematically well-defined machines,
Turing was able to capture the essence of computation
processes and algorithms.  There are a few other models of
computation~\cite{computerscience} but they all, except possibly the
quantum model to be considered later on, can be shown to be equivalent
to that given by Turing machines.  

Turing machines are equipped with an infinite one-dimensional tape over
which a head can move in discrete single steps backward and forward.
What symbol is already on the tape directly beneath the head and
what state the head is in will specify the new symbol (including the blank symbol)
to be written, if at all, over the old one and the new state of the head and
the direction in which the head moves next.  There needs to be only a
finite number of such symbols, similar to the finiteness of the number of letters in the
English alphabet.  The number of available states for the head is also
finite.  Everything about the machine is actually finite despite the appearance of the
infinite tape -- of which only a finite, but unbounded, portion will 
ever be used.  The infiniteness requirement for the tape is only there to enable
the accommodation of unbounded but finite sequences of ``words" and
``sentences" which can be meaningfully made out from the symbol
alphabets.

Initially, a finite portion of an otherwise blank tape is prepared with
some suitably encoded input for the Turing machine.
% , and off the machine goes with its head
% start over the input portion.  
Each Turing machine can do only one
specific task as completely defined by the transition function $T$:
\begin{eqnarray}
\langle {\mbox {\rm new symbol, new state, next movement}} \rangle &=& T({\mbox{\rm present
symbol, present state}}).
\end{eqnarray}
Once the task is done, the machine goes into a special halt state and
stops with the output written out on the tape ready to be inspected.
Transition functions are but representations for computer programs
implementing algorithms in general; so we will use interchangeably the names 
Turing machines and programs.

Such deceptively simple machines are so powerful that they can in fact capture the notion of
computation by algorithms as we intuitively know it.  
Turing went further to introduce his universal
machines, which are not restricted to do just a single computational task but 
can be ``reprogrammed" to do many different ones.   The key conceptual
leap that enables Turing universal machines is the recognition that the 
descriptions of Turing machines, given by the transition
functions, are not at all different to and can be encoded in the same way as the 
machines' very own inputs. 
The input to a Turing machine and the description of the Turing machine itself can be
put together to form a new, single input to yet another machine, a Turing universal machine.  
This latter machine only needs to emulate the encoded machine, which can be any well-formed
Turing machine, and thus acquires the status of being universal.
Universal machines are not unique and one can be encoded into the form
of an input for another as well.

The ability for a machine to act on its own kind enables
us to investigate the capability and limitations of Turing machines
using no other instruments other than the machines themselves.  It is
precisely this self-referential property that G\"odel exploited to embed
statements {\it about} arithmetics in statements {\it of} arithmetics in
his famous Incompleteness Theorem~\cite{godel}.  The embedding in the
Theorem is the same as the encoding of Turing machines into
input forms acceptable for universal machines and is achieved by converting 
the finite description of a Turing machine into a unique non-negative integer which
can then be expressed in binary or decimal or any other convenient notation.
The conversion is possible as we are only dealing here with machines having a
finite number of states, a finite number of symbols in its
alphabet, and only a finite number of movements for their heads.
Multi-tape Turing machines have been introduced and can simplify certain
computation; so can non-deterministic Turing machines whose next computation step at a
given instant can be chosen, with certain probabilities, among a finite
number of possibilities.  Nevertheless, these generalisations cannot
compute what a traditional Turing machine cannot do:  all of the Turing
machines have the same power in terms of computability~\cite{computerscience}.

Viewed from this perspective, (universal) Turing machines are just
functions from non-negative integers (encoding the input) to 
non-negative integers (encoding the
output).  But simple arguments can show that the class of functions
realised by Turing machines cannot be the same as the whole class of
functions from the set of natural numbers to the same set.  On the one
hand, the number of Turing machines is only countably infinite because each
machine can be mapped into a unique integer.  On the other hand, the
whole class of functions from natural numbers to natural numbers can be
shown by the Cantor diagonal arguments~\cite{recursive} to be 
uncountably infinite (in fact, this class of functions has the same 
cardinality as the set of
reals).  A pictorial and heuristic way to visualise this fact is 
where each mapping from natural numbers to themselves is
represented by a real-valued angle between the two semi-infinite half-lines, 
one of which contains the domain of the function, the other the
range.  There are as ``many" functions as there are angles between the
two lines.

The set of functions that are captured by and identified with Turing 
machines has been identified as that of so-called partial recursive functions 
and is briefly reviewed in the next Section.  From here on, we also 
call Turing machines interchangeably with programs and partial recursive functions.

\subsection{Recursive functions}
%basic primitive recursive functions
%composition
%premitive recursions
%premitive recursive functions (total)
%But there exists computable functions which are not premitive recursive
%--> $\mu$-recursive,$\mu$ unbounded minimisation, partial recursive
%functions

We begin with the set of {\em basic primitive recursive functions}, where all
the variables are in the set of non-negative integer $\cal N$,
\begin{enumerate}
\item The successor function: $s(x) = x + 1$;
\item The zero function: $z(x) = 0$;
\item The projections functions: $p^{(n)}_i(x_1, \cdots, x_n) = x_i,$
$1\le i \le n$.
\end{enumerate}
These are simple and intuitive functions.

From this basic set, other {\em primitive recursive functions} 
can be built using a finite number of the following operations
\begin{enumerate}
\item Composition: $f(x_1,\cdots,x_k) =
h(g_1(x_1,\cdots,x_k),\cdots,g_n(x_1,\cdots,x_k))$; where $h$ is an $n$-
variable function and the $n$ functions $g_i$ are $k$-variable.
\item Primitive recursion: Let $g$ and $h$ be functions with $n$ and
$n+2$ variables, respectively.  The $n+1$-variable function $f$ can then
be defined as 
\begin{eqnarray}
f(x_1,\cdots,x_n,0) &=& g(x_1,\cdots,x_n);\nonumber\\
f(x_1,\cdots,x_n,y+1) &=& h(x_1,\cdots,x_n,y,f(x_1,\cdots,x_n,y)).
\label{recursive}
\end{eqnarray}
\end{enumerate}
It is easy to see that the primitive recursive functions so constructed
are {\it total} in the sense that they are defined for {\em all} the 
non-negative integers.  

The set of primitive recursive functions is quite large but still 
inadequate and not large enough to encompass all the functions that are
admissible and computable by the Turing machines.  For this, we have to 
introduce yet another necessary operation, in addition to the two above,  
the unbounded $\mu$-minimisation which is defined as:
\begin{eqnarray}
f(x_1,\cdots,x_n) &=& \mbox{the least $y$ such that } p(x_1,\cdots,x_n,y) =
0;\nonumber\\
&\equiv& \mu y[p(x_1,\cdots,x_n,y)];\label{mu}
\end{eqnarray}
where $p(x_1,\cdots,x_n,y)$ is a total function for all the non-negative
integral values of the variables.  The general, mechanical method to
find $y$ that is applicable to all total functions $p$ is to step $y$
through 0, 1, 2, $\cdots$ until $p=0$ is satisfied, if this can indeed be satisfied.  
(For a particular
function $p$ one might be able to find $y$ in a different and quicker
way, but the mechanical method is applicable in general without any
privy knowledge about characteristics of $p$, as long as it is total.)

All these operations acting, in a finite number 
of steps, on the basic functions define the set of {\em partial recursive functions} which 
contains the set of primitive recursive functions as a proper subset.
The name {\em partial recursive} signifies the fact that there is no
guarantee that a (least) $y$ can always be found to satisfy the condition 
$p(x_1,\cdots,x_n,y) = 0$ for given $x_1,\cdots,x_n$ in the (unbounded)
$\mu$-minimisation.  If such a $y$ cannot be found, $f$ in~(\ref{mu}) is
undefined at that point.  The problem to establish whether $f$
is defined or not at a general point $(x_1,\cdots,x_n)$ in its
arguments, i.e. whether such a $y$ can be found, is a decision
problem which could have a binary answer, yes or no.  However, this 
particular problem is 
a mathematically undecidable problem, in the sense 
that there exists no general algorithm that can always yield the
required answer for any partial recursive function.

We define the set of {\em $\mu$-recursive functions} being the subset of
partial recursive functions where there always exists a decision procedure for
the $\mu$-minimisation operation, if
needed, that is, there always exists a value for $y$.  This set is bigger than the 
set of primitive recursive functions.  And it can be proved that a 
function is $\mu$-recursive if and only if there exists a Turing machine
which can compute it and then halts, i.e., if and only if it is Turing
computable.

The Church-Turing thesis stipulates that all the functions that
are {\em effectively computable} are in fact Turing computable, and vice versa,
thus restricting the intuitive and informal notion of 
computability to the well-defined mechanical operations of Turing machines.  
This identification is a thesis and not a theorem, not even a conjecture, 
because it can never be proven right in the mathematical sense.  Nevertheless,
it can be shown to be wrong if a counter-example can be found in which
the computation steps are clearly and acceptably identified.  Since it
is testable, the thesis does not have the status of an axiom in
Mathematics either.  

We dispute this thesis by showing in a later Section that there exist 
computable functions, computable by executing well-defined quantum mechanical 
procedures in a finite manner, that are not Turing computable.

The evaluation of partial recursive functions, even though well defined
(through the three operations of composition, primitive recursion and
$\mu$-minimisation upon basic functions and their resultants), 
may require an infinite number of steps -- in contrast to the 
$\mu$-recursive functions whose evaluation requires only a finite number of steps.  
That is, partial recursive functions can be implemented on a 
Turing machine with well-defined execution steps but the machine could 
never halt in some cases.

One example of non-halting is the simplest program to
find a number that is not the sum of four square numbers~\cite{penrose}.
All that the program can do is to step through the natural numbers one-by-one
and test each of these numbers by direct substitution of square numbers in increasing
order of magnitudes.  When a number can be found which is not the sum of
four squares, the program prints out that number and halts.  But this is
a false hope.  Constructed in this
way, such a program never halts because there exists Lagrange's Theorem 
which confirms that all numbers can in fact be expressed as the sum of four squares!

Another example can be seen through the famous but unproven Goldbach's conjecture
that every even number greater than 2 can be written as a sum of two
primes.  The simplest program would be one which steps through the
even numbers one by one, and for each of these even numbers tests for all 
prime numbers (less than the even numbers).  If an even number is found which 
is not a sum of two primes, the program prints out that number and halts.  In other 
words, the program halts if and only if the Goldbach's conjecture is
false.  So far, no counter-example to Goldbach's conjecture has been found this
way.  However, we cannot rely on such a program simply because it may never halt.

There are many other important conjectures and problems in Mathematics which can be
resolved if we somehow have a way to tell whether simple Turing
machines/programs, directly constructed similar to those above, will halt
or not.  Thus is the all-important Turing halting problem:  
given a Turing machine, the question is whether there is a general algorithm
which is able to determine if the machine would halt with a specified input.  
As is well-known in the framework of classical computability, this halting 
problem is as undecidable as the question whether a partial recursive function is
defined at a given argument.

\subsection{The Turing halting problem}
The halting problem for Turing machines is a manifestation of undecidability:  
a Turing computation is equivalent to the evaluation of a partial recursive function, 
which is only defined for a subset of the integers.  As this domain is
classically undecidable, one cannot always tell in advance whether the Turing machine 
will halt (that is, whether the input is in the domain of the partial recursive function) 
or not (when the input is not in the domain).

A version of the proof of the unsolvability of the halting problem
based on the Cantor diagonal argument goes as follows.  The proof is by contradiction 
with the assumption of the existence of a computable halting function $h(p,i)$ which has 
two integer arguments - $p$ is the G\"odel encoded integer number for the 
algorithm and $i$ is its (encoded) integer input:
\begin{eqnarray}
h(p,i) &=& \left\{\begin{array}{ll}
0 & \mbox{if $p$ halts on input $i$}\\
1 & \mbox{if $p$ does not}
\end{array} \right.  \label{h}
\end{eqnarray}
One can then construct a program $r(n)$ having one integer argument $n$ in
such a way that it calls the function $h(n,n)$ as a subroutine and 
\begin{eqnarray}
\left\{
\begin{array}{l}
\mbox{$r(n)$ halts if $h(n,n) = 1$} \nonumber\\
\mbox{$r(n)$ loops infinitely (i.e., never stops) otherwise.}
\nonumber
\end{array}
\right.
\end{eqnarray}
The application of the halting function $h$ on the program $r$ and input $n$ 
results in
\begin{eqnarray}
h(r,n) &=& \left\{\begin{array}{ll} 
0 & \mbox{if $h(n,n)=1$}\\
1 & \mbox{if $h(n,n)=0$}
\end{array}\right. \label{contradiction}
\end{eqnarray}
A contradiction is clearly manifest once we put $n=r$ in the last equation
above.  

The construction of such a program $r$ is transparently
possible, unless the existence of a computable $h$ is wrongly assumed.  
Thus the contradiction discounts the assumption that there is a classically 
algorithmic way to determine whether any arbitrarily given program with 
arbitrary input will halt or not.

However, this contradiction argument might be avoided if we distinguish and
separate the two classes of quantum and classical algorithms.  A {\it
quantum} function $qh(p,i)$, similar to eq.~(\ref{h}), can conceivably 
exist to determine whether any classical program $p$ will halt on any 
classical input $i$ or not.  The contradiction in eq.~(\ref{contradiction}) 
would be avoided if the quantum halting $qh$ cannot take as an argument the 
modified program $r$, which is now of {\it quantum} character because it now
has quantum $qh$ as a subroutine.  This will be the case if $qh$
can only accept integers while quantum algorithms, 
with proper definitions, cannot in general be themselves encoded as integers.  
% It is clear even in the case of a single qubit
% that the state $\alpha|0\rangle + \beta|1\rangle$ cannot be encoded as integers 
% for all $\alpha$ and $\beta$ - simply because of different cardinalities.  
In fact, the no-cloning theorem~\cite{no-clone} of quantum mechanics does 
restrict the type of operations available to quantum algorithms.

In essence, the way we will break the self-referential reasoning here by the
differentiation between quantum and classical algorithms is similar to
the way Bertrand Russell resolved the set theory paradox (to do with
``The set of all sets which are not members of themselves") by the
introduction of classes as distinct from sets.  (For other lines of
arguments, see~\cite{svozil, stannett, nz} for example.)

To investigate the decidability of the Turing halting problem in the framework 
of quantum computability, we will need to isolate the point which causes the
classical undecidability.  With this aim we turn to the Diophantine
equations in the next Section.

\subsection{Relation to Diophantine equations and Hilbert's tenth problem}
Identities between polynomials with integer coefficients in several
unknowns over the natural numbers have been studied for some time in 
mathematics under the name of Diophantine equations.
At the turn of the 
last century, David Hilbert listed, as challenges for 
the new century,  23 important problems among which the problem number 
ten is the only decision problem and could be rephrased as:
\begin{quote}
\it
Given any polynomial equation with any number of unknowns and with integer
coefficients (that is, any Diophantine equation):  To devise a universal 
process according to which it can be determined by a finite number of 
operations whether the equation has integer solutions.
\end{quote}

There are many important and interesting mathematical conjectures 
which can be proved or disproved depending on whether some corresponding
Diophantine equations have an integer solution or not.  These are, for
instances, Goldbach's conjecture which we already mentioned, the
Riemann hypothesis about the positions in the complex plane of the
zeroes of the zeta function, the four-colour conjecture for planar maps, 
Fermat's last theorem, etc
...  Thus, we can appreciate the importance of having a general method for all 
Diophantine equations, instead of considering each of them individually
on its own merits.

However, there are only few special cases of Diophantine equations that 
are solvable.  These include linear (first-degree) equations in the unknowns, 
of which the existence and absence of solutions can be inferred from the
Euclid's algorithm.  Also solvable are second-degree equations with
only two unknowns, that is, in quadratic forms. But Hilbert asked
for a {\em single} general and finite decision procedure that is applicable 
for any Diophantine equations.  
Little was it known that this problem is ultimately connected to the
seemingly unrelated notions of computation and Turing machines which
were to be introduced some 40 years later.  See~\cite{sciam} for a 
general introduction to Diophantine equations and Hilbert tenth problem.

Eventually, the Hilbert's tenth problem was finally shown to be 
undecidable in 1970 through a crucial step by Matiyasevich~\cite{hilbert, davis}:
The Hilbert's tenth problem could be solved if and only if the 
Turing halting problem could also be solved.  The two are simply equivalent.  
Consequently, as we have a proof in the last Section that the Turing's is not 
solvable, the Hilbert's tenth problem 
is noncomputable/undecidable in the most general sense if one accepts, as 
almost everyone does, the Church-Turing thesis of computability.
One would thus have to be content with the fact that individual Diophantine 
equations need to be considered separately, and may or may not be solved with 
a different approach anew each time.

For a precise discussion and the history of the negative result, 
see~\cite{hilbert, davis}; for a semi-popular account, see~\cite{casti}.  
We briefly sketch below some key arguments leading to the undecidable 
result.

It is recognised that Turing machines as devices mapping inputs to  
outputs are equivalent to partial recursive functions, whose domain are 
restricted to proper subsets of $\cal N$.  Without loss of generality, 
we can always restrict the computation of Turing 
machines to that of numeric partial recursive functions -- because for 
non-numeric algorithms there always exist mappings into the computation of some
corresponding partial recursive functions.  The question whether a
Turing machine would halt or not upon some particular input is now
equivalent to the question whether the corresponding argument value is in the domain
of the corresponding function or not.  When a machine cannot halt with a
given input is when the partial recursive function
corresponding to that machine is not defined at that particular
argument.  

Now, there is a universal representation, known as the Kleene normal form, 
for any partial recursive function $f$
\begin{eqnarray}
f(x_1,\cdots,x_n) &=& \psi(\mu y[\tau(g, x_1, \cdots, x_n, y)]),
\label{kleene}
\end{eqnarray}
where $\psi$ and $\tau$  are {\em fixed} primitive recursive functions, 
independent of the particular function $f$ on the left hand side.  The 
information about $f$ on the right hand side of~(\ref{kleene}) is encoded 
in the first argument of the function $\tau$ as the G\"odel number $g$ 
of $f$.  (We recall that the class of partial recursive functions is 
countably infinite and each function can be systematically given a unique 
identification number according to some G\"odel numbering scheme.)
It is the $\mu$-minimisation operation in~(\ref{kleene}) that could
potentially turn a total primitive recursive function, which is defined everywhere,
into a partial recursive function.  Thus, given 
$(x_1,\cdots, x_n)$ if there exists a (least) $y$ that for some integer $g$ solves 
$\tau(g, x_1, \cdots, x_n, y) = 0$ then the Turing machine, which corresponds to $g$, 
would halt upon the input corresponding to $(x_1,\cdots, x_n)$.  If no 
such $y$ exists, the Turing machine in question would not halt.

G\"odel, as quoted in~\cite{peter}, has shown that the $\mu$-minimising
operation~(\ref{mu})
can always be represented as some arithmetic statement with a set of identities 
between multi-variate polynomials over the integers, together with a finite number 
of the existential quantifiers, $\exists$, and bounded versions of the universal 
quantifiers, $\forall$.  
(The boundedness comes from the requirement of some {\em least} number
that satisfies a $\mu$-minimisation as defined in~(\ref{mu}).)
Note that quantifiers 
hold a crucial role in arithmetic.  Restricted arithmetics with only
the addition operation or the multiplication operation, but not both, are in
fact complete:  any of their statements can be decided to be provable or
not within the relevant frameworks.  On the other hand, we have the famous G\"odel's 
Incompleteness Theorem for `ordinary' arithmetic which has both addition and 
multiplication operations. Normally we would think that multiplication is 
just a compact way to express long and repeated additions, and might thus be
puzzled over the difference in completeness.  But it is the quantifiers 
that make such a huge difference.

Arithmetic statements for solutions of Diophantine equations 
(i.e. the Hilbert's tenth problem) can only involve existential quantifiers, 
but the elimination of bounded universal quantifiers in the $\mu$-minimisation
necessitates the appearance of exponentiation of variables.  A famous
example of variable exponentiation is the equation of Fermat's
last theorem:
\begin{eqnarray}
(x+1)^{(u+3)} + (y+1)^{(u+3)} + (z+1)^{(u+3)} &=& 0,
\label{fermat}
\end{eqnarray}
which has no non-negative integer solutions for the unknowns $x$, $y$, $z$ 
and $u$.  

After various attempts by many people, it was finally shown that exponentiation 
is indeed Diophantine -- that is, variable exponents can be eliminated to result in
only polynomial Diophantine equations.  This proves the equivalence between the two 
decision problems and implies that the Hilbert's tenth is not computable in the Turing 
scheme of computation.

% $\mu$-minimisation is Diophantine?  Perhaps yes, direct construction 
% (adopts Peter's page 265, section 4, or with the semi-closure of being
% Diophantine -- Davis's.  But need the proofs that bounded universal
% quantifier can be eliminated and exponential Diophantine is Diophantine) 
% or via (Recursive enumerable sets are identical to those of Diophantine; or every
% Turing semi-decidable set is Diophantine.)
% That is, every Turing machine (program/algorithm) is representable by a 
% particular Diophantine equation, and that Turing machine halts upon 
% some particular input iff that Diophantine equation has solutions. (eq 5.5.4
% of Matiyasevich's.)

%equivalent to Turing computable:  computable in a finite number of well
%defined steps using finite physical means -- algorithms <-- definition

There are many other noncomputable problems, some are also equivalent to the
Turing halting problem, and others belong to altogether different
non-computability classes.

\subsection{Some other noncomputables}\label{sec:others}
Similarly based on the Turing machines, many other problems 
can be shown to be equivalent to the Turing halting problem, and thus
noncomputable.  We can name in this class~\cite{computerscience} the tiling problem 
in a plane, Post's correspondence problem, Thue's word problem, Wang's 
domino problem, etc ...  For example, the tiling problem asks for a
decision procedure to see if any given finite number of sets of tiles 
(but each set has an infinite number of the same tiles available) can tile the 
first quadrant of the plane or not, with a specified tile at the origin at the
lower left corner and some adjacency rules for the tiling.

One could also ask questions related to and generalising the Hilbert's
tenth problem such as~\cite{davis}: Is there a single algorithm for testing 
whether the number of solutions for any Diophantine equation is finite, or is
infinite, or equal to one, or is even?

Besides the above equivalent to the Turing halting problem, there are yet
others which belong to different classes in the non-computability
hierarchy.  The most famous of all must be the G\"odel's
Incompleteness Theorem.  It establishes that any finitely axiomatic,
consistent mathematical system sufficiently complex to embrace
Arithmetic must be incomplete -- that is, there exist some statements
whose truth cannot be confirmed or denied from within the system,
resulting in undecidability.  (Interestingly, it is 
precisely the statement about consistency of the system that is 
neither provable nor deniable.)  We can see that this undecidability is more
than that of Turing machines, whose applicable mathematical 
statements can only have bounded universal quantifiers as opposed to
arbitrarily unbounded universal quantifiers of Arithmetic at large.

Chaitin, approaching from the perspectives of Algorithmic
Information Theory~\cite{chaitin2, chaitin3}, has shown that there exist many
unprovable statements in Arithmetic simply because they have irreducible 
algorithmic contents, measurable in bits, that are more than the complexity, 
also measurable in bits, of the finite set of axioms and inference rules 
of the system.  Infinitely irreducible algorithmic complexity is randomness 
which exists even in pure mathematics.
And more frustratingly, we can never prove randomness since we could 
only ever deal with finite axiomatic complexity.  Chaitin, to illustrate 
the point, has introduced the number $\Omega$ as
the halting probability for a random program, with some random input, 
being emulated by a particular Turing machine.
$\Omega$ is an average measure over all programs run on the universal Turing 
machine.
This number has many interesting properties, and has been generalised to a 
quantum version~\cite{svozil}, but we only mention here the linkage 
between this number and polynomial Diophantine 
equations.  When expressed in binary, the value of the $k$-th bit of 
$\Omega$ is respectively 0 or 1 depending on whether some Diophantine equation, 
corresponding to the Turing machine in consideration,
\begin{eqnarray}
C(k,N,x_1,\cdots,x_K) &=& 0,
\label{omeg}
\end{eqnarray}
for a given $k>0$, has solutions in non-negative integers 
$(x_1, ..., x_K)$ for
finitely or infinitely many values of the parameter $N>0$.

From this Diophantine representation we can see that the noncomputability 
of $\Omega$ is ``more" than the noncomputability of the Turing halting problem.
Even if, somehow, we have an algorithm for the latter to decide the
equations in~(\ref{omeg}) for each $N$ respectively, we would still need to 
apply the algorithm for an infinite number of times just to get a single
digit for $\Omega$!  An infinite number of times cannot be normally performed in a finite 
time, hence the different class of noncomputability from that of 
the Turing halting problem.
We will come back to the $\Omega$ number with a hypothetical algorithm 
in Quantum Field Theory later.

We end our brief review of relevant mathematical concepts here with a
discussion of computable numbers.  If we had restricted such numbers to only
those that can be output by some Turing machine (which then halts), then
we would have had to effectively restrict ourselves to integers and to
treat irrational numbers, like $\sqrt 2$ and $\pi$,
as noncomputable.  This is clearly undesirable.  A better and more practical
definition of computable numbers are those which we can approximate to
an arbitrary degree of accuracy with (integer/rational) outputs of some Turing machines.
This
definition of computability of a number when it can be evaluated to any
degree of accuracy is sufficient to interpret the number and to establish 
its relationship with other numbers.  And that all we need in computing
any number: its place in relation to other numbers.

Note that, in this sense, the Chaitin's $\Omega$ is not computable because we
cannot estimate this number to an arbitrarily high accuracy, although it
is bounded from above by unity (being the maximum probability) and can be
approximated only from below by some very slowly converging process~\cite{chaitin2} 
whose convergence rate is indeterminable.

\section{Quantum Mechanics}\label{sec:qm}
\subsection{The postulates of measurement and associated problems}\label{subsec:measure}
% probability, measurables
Quantum Physics, including Quantum Mechanics and Quantum Field Theory,
is the most successful theory that we have in Science for the description and
prediction of phenomena in Nature.  And so far there is not a single
discrepancy between the theory and experiments.  

According to Quantum Mechanics, pure states of a physical system can capture the 
most that can be said about the system and 
are associated with vectors, unique up to phases, in some abstract linear vector 
Hilbert space.  When  the system, particularly when it is a subsystem of 
a bigger entity, cannot be described by a pure state but 
is in a mixed state, the language of density matrices would be necessary 
for its description.   Acting on the Hilbert space are linear operators,
of which hermitean and unitary operators are of particular interest.

In the Schr\"odinger picture where the time dependency is explicitly
carried by the states, the time evolution of the system is governed by
the Schr\"odinger equation, in which the hermitean Hamiltonian operators
play a unique role.  In general, each physical observable is associated
with a hermitean operator; the Hamiltonian operator, for instance, is
associated with the system's energy.  The real-valued eigenvalues (which can be 
continuous or discrete) of these hermitean operators restrict the obtainable
values under observation.  Each time when the associated observable is measured, only one 
single value, among the eigenvalues given, is obtained.  Repetitions of
the measurement under identical conditions could yield different measured
values each time.  And the probability of
getting a particular eigenvalue in a measurement is given by the square
of the absolute value of the inner product between the corresponding
eigenvector and the state describing the system at that instant.  If
the system is in a mixed state described by a density matrix, the
probability is then given by the trace of the product between the
density matrix and the corresponding projector associated with the
eigenvector in question.

After a measurement, the state of the system is a pure state whose
representing vector is the same as the eigenvector, up to a phase, 
corresponding to the eigenvalue obtained.  Note that measurement thus is a 
non-unitary and irreversible operation in general, unless the system 
is already in the observable eigenstate.  Different observables can be measured 
simultaneously, with the same degree of statistical accuracy, 
only when the associated hermitean operators commute with each other.

But already seeded in the summary of quantum mechanical postulates above 
is a fundamental problem of inconsistency.  The act of measurement,  
on the one hand, is a process unfolded itself in time.  On the other hand, why
should it not be governed by the unitary Schr\"odinger time-evolution
operator?  

Even when we extend the system to a larger system consisting of
the considered system and the measuring apparatus, we still face the same
Quantum Measurement Problem, under a slightly different guise.
In setting up a measurement we effectively establish some
correlation between the state of the measured system and the pointers of
the measuring apparatus.  This correlation is one-to-one for the
eigenstates of the operator representing the measured observable: if the
system is in the state $|e_i\rangle$, $i=1,2$, 
then the pointer, initially in a neutral state $|A_0\rangle$, is subsequently in
the state $|A_i\rangle$ respectively, 
\begin{eqnarray}
|e_i\rangle|A_0\rangle \to |e_i\rangle|A_i\rangle.
\end{eqnarray}
Now, according to the quantum mechanical principles of superposition and linearity, it
follows that if the system is in a superposition then so is the pointer, 
\begin{eqnarray}
\left(\alpha |e_1\rangle + \beta |e_2\rangle\right) |A_0\rangle \to \alpha |e_1\rangle |A_1\rangle 
+ \beta |e_2\rangle|A_2\rangle.
\label{measure}
\end{eqnarray}
The problem is that we have never been able to observe the classical
pointer in an entangled state as in the right hand side of the last
expression.  Instead, we either get the pointer position $A_1$ with a probability 
proportional to $|\alpha|^2$ or $A_2$ with a probability proportional to
$|\beta|^2$.  

To resolve this problem one could postulate, as in the
Copenhagen interpretation, {\it some} undefined separation
between the quantum and the classical worlds, or modify the theory to have elements of
non-linearity admitted.  (This second situation is unlike mathematics in the way 
that the postulates/axioms of
a physical theory can themselves also be subjects of investigation.)
But if we believe in Quantum Mechanics as the universal and 
fundamental theory then we have a difficult problem at hand.  
Various resolution attempts ranging from the many-world interpretation to 
decoherence have not been deemed successful.

We shall leave the measurement problem here, and only wish to emphasise that
the power of all quantum algorithms in quantum computation
relies crucially on such mysterious measurement processes.

Also for later use, we now introduce the concept of {\em measurable}
quantities~\cite{geroch}.  Analogous to the concept of computable
numbers discussed in Sec.~\ref{sec:others}, a number $w$ is deemed measurable if
there exists a finite set of instructions for performing an experiment
such that a technician, given an abundance of unprepared raw materials
and an allowed error $\epsilon$, is able to obtain a rational number
within $\epsilon$ of $w$.  The technician is analogous to the computer,
the instructions analogous to the computer program, the ``abundance of unprepared raw
materials" analogous to the infinite Turing tape, initially blank.

In particular, not only the (stochastic) outcomes of an observable are 
obviously measurable and of interest but so are the probability distributions 
for these outcomes.  The probabilities can be obtained within any given accuracy by 
increasing the number of measurement repetitions.  Later, we make full use of 
this crucial fact that {\em the quantum mechanical probabilities are 
% both computable from the theory of Quantum Mechanics 
and measurable in principle.}

\subsection{Peculiarities of Quantum Mechanics}
One important, but least understood, property of Quantum Mechanics is the randomness in 
the outcome of a quantum measurement.  Even if we prepare the initial
quantum states to be {\em exactly} the same in principle, 
we can still have 
different and random outcomes in subsequent measurements.  
Such randomness is a fact of life in the quantum reality of
our universe.

To reflect that intrinsic and inevitable randomness, the best that Quantum
Mechanics, as a physical theory of nature, can do is to list, given the 
initial conditions, the possible values for measured quantities and the
probability distributions for those values.  Both the values and the
probability distributions are computable in the sense discussed at the end of
Sec.~\ref{sec:others} for Hilbert spaces with finite dimensions.  With
countably infinite dimensions, Quantum Mechanics can be {\em
noncomputable} unless we can control the accuracy of calculated values
by suitable truncation of the infinite dimensional space.  (The
mathematical noncomputability of the Hilbert's tenth problem is reflected through
this infinite dimensions of the Fock space in our formulation of the
problem.  But there is a way out as discussed below.)

On the other hand, not only the values registered in the measurement of
some measurable but also the associated probability distributions
are measurable in the sense that they can be obtained to any
desirable accuracy by the act of physical measurements.  
Normally, the
values for measurable are quantised so they can be obtained exactly; the
probability distributions are real numbers but can be obtained to any
given accuracy by repeating the measurements again and again 
(each time from the same initial quantum state) until the 
desired statistics can be reached.  That is how the computable numbers 
from Quantum Mechanics can be judged
against the measurable numbers obtained from physical experiments.  Thus far, 
there is no evidence of any discrepancy between theory and experiments.

% QM measurables' values and probability: measurable and computable.

Randomness is, by mathematical definition, incompressible and irreducible.  
In Algorithmic Information Theory, Chaitin~\cite{chaitin2} defines 
randomness by program-size complexity: a binary string is considered 
random when the size of the shortest program that generates that string 
is not ``smaller", as measured in bits, than the size of the string itself.  We refer the readers 
to the original literature for more technically precise definitions for 
the cases of finite and infinite strings.

Paradoxically, the quantum reality of Nature somehow allows us 
to {\em compress} the {\em infinitely incompressible} randomness into the 
{\em apparently finite} act of preparing the same quantum state over and over again
for subsequent measurements!  This quantum mechanically implied infinity
seems to be both needed for and consistent
with the finitely measured, see~\cite{kieurandom} and references therein for further
discussion.

\subsection{Coherent states}
One of the simplest and most widely applicable problems in Quantum Mechanics is that of the 
(one-dimensional) Simple Harmonic Oscillator (SHO) with the Hamiltonian 
\begin{eqnarray}
H_{\rm SHO} &=& (P^2 + X^2)/2,
\label{sho}
\end{eqnarray}
which can also be expressed as
\begin{eqnarray}
H_{\rm SHO} &=& a^\dagger a + \frac{1}{2}.
\label{sho2}
\end{eqnarray}
The operators $a^\dagger$, $a$ are linearly related to the position 
and momentum operators, which satisfy the commutation relation $[P, X] = i$, 
\begin{eqnarray}
X &=& \frac{1}{\sqrt 2}(a + a^\dagger), \nonumber \\
P &=& \frac{i}{\sqrt 2}(a - a^\dagger).
\label{ca}
\end{eqnarray}
The operators $a^\dagger$, $a$ thus satisfy different commutation relations
\begin{eqnarray}
&&[a, a^\dagger] = 1 ,\nonumber \\ 
&&[a, a] = [a^\dagger, a^\dagger] = 0.
\end{eqnarray}

The spectrum of the number operator $N=a^\dagger a$ that appears
in~(\ref{sho2}) is discrete and spans over the natural numbers.  Its
eigenstates are termed the number states $|n\rangle$,
\begin{eqnarray}
N |n\rangle &=& n|n\rangle; \;\;\; n=0,1,2,\cdots
\label{number}
\end{eqnarray}
These eigenstates also constitute an orthonormal basis for a 
Fock space, a special type of Hilbert space,
% \begin{eqnarray}
% \langle m | n\rangle &=& \delta_{mn}, 
% \end{eqnarray}
and can be constructed by the operators $a^\dagger$ acting on the special
``vacuum" state $|0\rangle$, the lowest-eigenvalue state,
\begin{eqnarray}
|n\rangle &=& \frac{a^{\dagger n}}{\sqrt n!} |0\rangle,
\end{eqnarray}
from which follow the recursive relations
\begin{eqnarray}
a^\dagger |n\rangle &=& \sqrt{n+1} |n+1\rangle,\nonumber\\
a |n\rangle &=& \sqrt{n} |n-1\rangle.
\end{eqnarray}
These relations lead us to the names {\em creation} and {\em annihilation} 
operators respectively for $a^\dagger$ and $a$.

The number state $|n\rangle$ can be realised in Quantum Optics as one 
having a definite number of $n$ photons, all at the same frequency.  
But these number states are not
the states of travelling optical modes generated by idealised 
lasers~\cite{enk}, which have an indefinite number of photons.  For the
description of these modes, we need the coherent states~\cite{walls},
$|\alpha\rangle$, which are the eigenstates of $a$ and are labeled 
by the complex number $\alpha$,
\begin{eqnarray}
a |\alpha\rangle &=& \alpha |\alpha\rangle.
\end{eqnarray}
With the relation to the number states,
\begin{eqnarray}
|\alpha\rangle &=& {\rm e}^{-\frac{|\alpha|^2}{2}}\sum_{n=0}^{\infty} 
\frac{\alpha^n}{\sqrt n!} |n\rangle, \nonumber\\
&=& {\rm e}^{-\frac{|\alpha|^2}{2}} {\rm e}^{\alpha a^\dagger} |0\rangle,
\end{eqnarray}
the coherent states are not orthogonal but can still be used for
spanning the Hilbert space.  
They have some unique and nice properties, one of which is that they are 
the states that optimise the amplitude-phase Heisenberg uncertainty relation. 
The other fact, which we will exploit later, is that they are also
the ground states, i.e. the eigenstates having the lowest ``energy"
eigenvalues, for the family of semi-definite Hamiltonians 
\begin{eqnarray}
H_{\alpha} &=& (a^\dagger - \alpha^*)(a - \alpha).
\label{alpha}
\end{eqnarray}
With the simple substitution $a_\alpha = a - \alpha$ we are back to a 
family of $\alpha$-labeled SHO Hamiltonians~(\ref{sho}), except for the additive
constant, all of which have the same spectrum over the natural numbers but with 
different sets of eigenvectors $|n_\alpha\rangle$.

\subsection{The quantum adiabatic theorem}\label{sec:adia}
The dynamical evolution of a quantum system is governed by the Hamiltonian
through the Schr\"odinger equation.  If the system is closed then the
Hamiltonian is time independent.  If the system is subject to external
influences, whose dynamics are not of direct concern to the investigation,
then the Hamiltonian is time dependent; and the modification
in the quantum state of the system critically depends on the time $T$
during which the change of the Hamiltonian takes place.  This dependency
is particularly simplified when the rate of change of the external fields 
is very fast compared to some intrinsic time scale, whence we can apply the 
sudden approximation, or is very slow, whence we can appeal to the quantum
adiabatic theorem.

The sudden approximation says that if the time change $T$ is sufficiently
fast relative to the inverse of the average Hamiltonian during
that time, $\Delta \bar {H}$,
\begin{eqnarray}
T &\ll& \hbar/\Delta \bar {H},
\label{sudden}
\end{eqnarray}
then the dynamical state of the system remains essentially unmodified.

On the other hand, in the case of an infinitely slow, or adiabatic
passage, if the system is initially in an eigenstate of the
Hamiltonian at the initial time it will, under certain conditions, pass
into the eigenstate of the Hamiltonian at the final time, that derives
from it by continuity~\cite{messiah}.  This is the content of the
adiabatic theorem, provided the following conditions are satisfied
throughout the relevant time interval:
\begin{itemize}
\item The instantaneous eigenvalues remain distinct;
\item The first and second derivatives of the instantaneous
eigenvectors with respect to time are well-defined and 
piece-wise continuous.
\end{itemize}

This important theorem has been exploited for a model of
quantum computation involving the ground state and to be discussed 
below.  In practice, the 
evolution time $T$ is finite but the more it satisfies the condition 
\begin{eqnarray}
T &\gg& \frac{\parallel\Delta H\parallel}{g^2},
\label{adiabatic}
\end{eqnarray}
then the higher the probability that the system remains in the
instantaneous eigenstate.  In the above,
\begin{eqnarray}
\parallel \Delta H \parallel &\equiv& \max_{0\le t \le T} \left|\langle 
e(t)|H(T) - H(0)| g(t)\rangle\right|,
\label{norm}
\end{eqnarray}
and
\begin{eqnarray}
g &\equiv& \min_{0\le t \le T} \left(E_e(t) - E_g(t)\right),
\label{gap}
\end{eqnarray}
where $|g(t)\rangle$ and $|e(t)\rangle$ are respectively the instantaneous
ground state and first excited state with
instantaneous eigenvalues $E_g(t)$, $E_e(t)$.  The critical scale is
this gap $g$.

\section{Quantum Computation}\label{sec:qc}
The underlying laws of all physical phenomena in Nature, to the best of our knowledge, 
are those given by quantum physics.  However, the best present day computers, as they are 
of classical nature, cannot even in principle simulate quantum systems efficiently. 
Feynman pointed out in 1982 that~\cite{feynman}, see also~\cite{benioff}, only quantum 
mechanical systems may be able to simulate other quantum systems more efficiently.
Furthermore, according to Moore's law, the exponential rate of miniaturisation of 
micro-electronic semiconductor devices will soon, if not already, take us to the sub-micron 
and nano dimensions and beyond.  On this scale, quantum physics will become more and more 
relevant in the design and production of computer components.  Heat dissipation in irreversible 
computation will be yet another problem at these dimensions.  Even though reversible 
classical computation can be implemented in principle, quantum computation, being almost 
reversible except the final read-out by measurement, could automatically minimise 
this heating problem.

Lastly, the notion of effective computability, as delimited by the Church-Turing
thesis ``Every function which would be naturally regarded as 
computable can be computed by the universal Turing machine", 
begs the question whether it could be extended with quantum principles.  
Initial efforts seem to confirm that
quantum computability is no more than classical and mathematical
computability~\cite{QTM}.  However, more recent
indications may prove otherwise~\cite{kieu, nz}.  We will
come back to this computability notion in a Section below.

All of the above have inevitably led to the recent convergence of quantum
physics, mathematics, and computing and information processing.

\subsection{``Standard" Model of Quantum Computation}
According to the ``standard'' model of quantum computation (see~\cite{qc} for instance),
which is a direct generalisation of
classical digital computing, the fundamental unit of a quantum computer is the quantum bit, 
shortened as {\it qubit}, which is the generalisation of a binary bit.  Physical implementation of 
a qubit could be any (measurable) two-state system: the up and down values of a quantum 
spin, or the two polarisation states of a photon, etc ... But unlike the binary bit, a qubit 
can be in a superposition state of its two states/values.  Upon measurement the superposition 
is destroyed, revealing one of the two classical values of a qubit.  The two states of a 
qubit, denoted by $|0\rangle$ and $|1\rangle$, should be unambiguously distinguishable 
by measurement and thus be orthogonal to each other.  

There are three stages of operation for a quantum computer, corresponding to the input, 
the processing, and finally the output.  The input preparation stage can
be and has been
carried out in laboratories for certain well-known systems.  So can the output 
stage in which the
output is read out by an act of measurement -- even though quantum measurement is not that 
well understood, as already alluded to in Sec.~(\ref{subsec:measure}).  
The speed of state preparation and measurement, 
which should be carried out 
in such a way as not to perturb other subsystems/qubits not being directly measured, is 
crucial for quantum computation.  The information processing stage is the most difficult to be 
implemented.  
In principle, it is governed by unitary evolution of a set of qubits well isolated from the 
surroundings to avoid as much of the decoherence effects of the environment as possible.  
The discovery of error correcting codes for quantum computation was a
pleasant surprise.  Without this possibility, realisation of quantum computation would have 
been unthinkable as computers are inevitably and constantly subject to errors induced by 
either the internal interactions or the environment or both. 

The power of a quantum computer firstly lies in the massive parallelism resulting directly 
from the superposition possibility of the quantum states.  If each qubit is a superposition 
of two states then the measurement of such a superimposed $N$-qubit system could in general 
access $2^N$ distinguished states simultaneously.  However, such quantum parallelism is 
not that useful because of the probabilistic nature of the measured outcomes.  (After all, 
many classical wave systems, like water waves, can also have superposition but cannot 
provide a better computation model.)  The second and most important power of quantum 
computation is thought to have its root in {\it quantum entanglement}~\cite{bell}, which 
has no counterpart in the classical world (even though it might be expensively simulated by 
classical means).  Quantum entanglement provides the extra dimensions in information storage 
and processing that distinguishes the quantum from the classical.  It is the entanglement that 
allows us to {\it control} the massive quantum parallelism through selective interference
of different computational paths to extract the information desired.  

%  progress so far in algorithms and other achievements
These characteristics have been exploited to reduce the computational complexity of some 
problems.  So far there are only few quantum algorithms discovered~\cite{shor, grover}; 
most notable is Shor's
factorisation algorithm which employs Quantum Fourier Transformation.  
QFT is the only known quantum algorithm that could offer an exponential increase in 
computational speed, due to the interference of different computation paths 
(as Fourier Transformation is intimately linked to interference) and to quantum 
entanglement.  More quantum algorithms are urgently needed.

The approach above with qubits and unitary gates of so-called 
quantum networks has been accepted as the standard model for quantum
computation.  It has been argued~\cite{QTM} that the
computability obtainable in this model is not better but is the same as
classical computability.  

However, it is not the only model available.  

\subsection{Quantum Adiabatic Computation}
Among the alternative models for quantum computation is the recent proposal~\cite{adiabatic} 
to employ quantum adiabatic processes for computation.    
The idea is to encode the solution of some problem to be solved into the
ground state, $|g\rangle$, of some suitable Hamiltonian, $H_P$.  
But as it is easier to implement the Hamiltonian
than to obtain the ground state, we should start the
computation in yet a different and readily obtainable initial
ground state, $|g_I\rangle$, of some initial 
Hamiltonian, $H_I$, then deform this Hamiltonian in a time $T$
into the Hamiltonian whose ground state is the desired one, through a time-dependent process,
\begin{eqnarray}
{\cal H}\left(\frac{t}{T}\right) &=& \left(1-\frac{t}{T}\right)H_I + \frac{t}{T}H_P.
\label{Hamiltonian}
\end{eqnarray}
The adiabatic theorem of Quantum Mechanics, Sec.~\ref{sec:adia},
stipulates that if the deformation time is sufficiently slow compared
to some intrinsic time scale, the initial state will evolve into the desired
ground state with high probability -- the longer the time, the higher
the probability.

\section{Quantum algorithm for the Hilbert's tenth
problem}\label{sec:10}
\subsection{General approach}
It suffices to consider non-negative solutions, if any, of a
Diophantine equation.  
Let us consider a particular example
\begin{eqnarray}
(x+1)^3 + (y+1)^3 - (z+1)^3 + cxyz = 0, && c\in Z,
\label{eq}
\end{eqnarray}
with unknowns $x$, $y$, and $z$.
%which is a particular case of the Fermat's last theorem.  
To find out
whether this equation has any non-negative integer solution by quantum
algorithms, it requires the realisation of a Fock space.
Upon this Hilbert space, we construct the Hamiltonian corresponding
to~(\ref{eq})
\begin{eqnarray}
H_P &=& \left((a^\dagger_x a_x+1)^3 + (a^\dagger_y a_y+1)^3 - (a^\dagger_z
a_z+1)^3 + c(a^\dagger_x a_x)(a^\dagger_y a_y)(a^\dagger_z a_z) \right)^2,
\nonumber
\end{eqnarray}
which has a spectrum bounded from below -- semidefinite, in fact.  

Note that the operators $N_j = a^\dagger_j a_j$ have only 
non-negative integer eigenvalues $n_j$, and that $[N_j, H_P] = 0 = 
[N_i, N_j]$ so these 
observables are simultaneously measurable.   The 
ground state $|g\rangle$ 
of the Hamiltonian so constructed has the properties
\begin{eqnarray}
N_j|g\rangle &=& n_j|g\rangle, \nonumber\\
H_P|g\rangle &=& \left((n_x+1)^3 + (n_y+1)^3 - (n_z+1)^3 + cn_xn_yn_z
\right)^2|g\rangle \equiv E_g |g\rangle,
\nonumber
\label{eigenvalues}
\end{eqnarray}
for some $(n_x,n_y,n_z)$.

Thus, a projective measurement of the energy $E_g$  of the ground state 
$|g\rangle$ will yield 
the answer for the decision problem: The Diophantine equation has at least one
integer solution if and only if $E_g = 0$, and has not otherwise.  (If
$c=0$ in our example, we know that $E_g > 0$ from Fermat's last theorem.)

If there is one unique solution then the projective
measurements of the observables corresponding to the operators
$N_j$ will reveal the values of various unknowns.  If there
are many solutions, finitely or infinitely as in the case of the Pythagoras
theorem, $x^2 + y^2 -
z^2 = 0$, the ground state $|g\rangle$ will be a linear superposition of 
states of the form $|n_x\rangle|n_y\rangle|n_z\rangle$, where
$(n_x,n_y,n_z)$ are the solutions. In such a situation, the measurement may 
not yield all the solutions.  However, finding all the solutions is not
the aim of a decision procedure for this kind of problem.  

Notwithstanding this, measurements of $N_j$ of the ground state would always 
yield some values $(n_x,n_y,n_z)$ and a straightforward substitution would 
confirm if the equation has a solution or not.  Thus the measurement on the 
ground state either of the energy (with respect to the hermitean operator $H_P$, 
provided the zero point can be calibrated)
or of the number operators will be sufficient 
to give the result for the decision problem.  

The quantum algorithm with the ground-state oracle is thus clear:
\begin{enumerate}
\item  Given a Diophantine equation with $K$ unknown $x$'s
\begin{eqnarray}
D(x_1,\cdots,x_K) &=& 0,
\end{eqnarray}
we need to simulate on some appropriate Fock space
the quantum Hamiltonian 
\begin{eqnarray}
H_P &=& \left(D(a^\dagger_1 a_1,\cdots, a^\dagger_K a_K) \right)^2.
\end{eqnarray}
\item  If the ground state could be obtained with high probability and 
unambiguously verified,
measurements of appropriate observables would provide the answer for
our decision problem.
\end{enumerate}
The key ingredients are the availability of
a countably infinite number of Fock states, the ability to construct/simulate
a suitable Hamiltonian and to obtain and verify its ground state.
As a counterpart of the semi-infinite tape of a Turing machine, the Fock 
space is employed here instead of the qubits of the more well-known
model of quantum computation.  Its advantage over the infinitely many 
qubits which would otherwise be required is obvious.  

One way to construct any suitable Hamiltonian so desired is through
the technique of ref.~\cite{continuous}.  We consider the hermitean 
operators, where $j$ 
is the index for the unknowns of the Diophantine equation,
\begin{eqnarray}
X_j &=& \frac{1}{\sqrt 2}(a_j + a^\dagger_j), \nonumber \\
P_j &=& \frac{i}{\sqrt 2}(a_j - a^\dagger_j),\\   \nonumber 
 [P_j, X_k] &=& i\delta_{jk}.
\end{eqnarray}
Together with the availability of the fundamental Hamiltonians 
\begin{eqnarray}
X_j, P_j, (X^2_j + P^2_j), \pm(X_kP_j + P_jX_k), \mbox{ and } (X^2_j + P^2_j)^2
\label{basic}
\end{eqnarray}
one could construct the unitary time evolutions corresponding to Hamiltonians
of arbitrary hermitean polynomials in $\{X_j,P_j\}$, and hence in $\{a^\dagger_j
a_j\}$, to an arbitrary degree of accuracy.  
These fundamental Hamiltonians correspond to, for examples, translations,
phase shifts, squeezers, beam splitters and Kerr
nonlinearity~~\cite{continuous}.

With the polynomial Hamiltonian constructed, we need to obtain its ground 
state.  Any approach that allows us to access the ground state will
suffice.  One way is perhaps to use that of quantum annealing or 
cooling~\cite{qannealing}.  Another way
is to employ the quantum computation method of 
evolution with time-dependent Hamiltonians.

\subsection{Time-dependent Hamiltonian approach}
Of the theory of Quantum Mechanics we cannot in general compute in infinite
dimensional Hilbert spaces the various observables, including the
probability distributions for these.  The reason is that we can only employ
some truncated versions of the Hilbert spaces but may not know, from
mathematical arguments solely, the relationship between the truncation and the
accuracy of those calculated from the theory is.  On the other hand, we
do have some access to the presumably infinite physical world; but the
access is quite limited, through the act of measurement only.  Our
approach is to combine the two, and exploit the measurability in the
physical world to compute the mathematically noncomputable.
Below is an algorithm based on the exploitation of both the
presumably infinite physical world and the theory of Quantum Mechanics
calculated in a finite manner on Turing machines.  The algorithm
presented may not be the most efficient; there could be many other
variations making better use of the same philosophy.

In the physical world, it is in general easier to implement some Hamiltonian
than to obtain its ground state.  We thus should start with 
yet a different and readily obtainable initial
ground state, $|g_I\rangle$, of some initial 
Hamiltonian, $H_I$, then deform this Hamiltonian in a time $T$
into the Hamiltonian whose ground state is the desired one, through a 
time-dependent process represented by the interpolating Hamiltonian ${\cal H}(t/T)$.

One could start, for example, with the Hamiltonian $H_I$, for some
$\alpha's$,
\begin{eqnarray}
H_I = \sum_{i=1}^K (a^\dagger_i -\alpha^*_i)(a_i - \alpha_i),
\label{init}
\end{eqnarray}
which admits the readily achievable coherent state 
$|g_I\rangle = |\alpha_1\cdots\alpha_K\rangle$ as the ground state.  
Then, one forms the slowly varying Hamiltonian
${\cal H}(t/T)$ in~(\ref{Hamiltonian}), which interpolates in the time interval 
$t\in[0,T]$ between the initial $H_I$ and $H_P$.  
\begin{itemize}
\item {\em Step 0:}  Choose an evolution time $T$, a probability $p$ which can be 
made arbitrarily closed to unity, and an accuracy $0<\epsilon<1$ which can be made 
arbitrarily small.
\item  {\em Step 1 (on the physical apparatus):}  Perform the {\it physical} 
quantum time-dependent process which is governed by the time-dependent Hamiltonian
${\cal H}(t/T)$ and terminates after a time $T$.   Then, by projective
measurement (either of the observable $H_p$ or the number operators $\{N_1, \ldots, N_K\}$) 
we obtain some state of the form $|\cdots n_i \cdots\rangle$, $i=1,\cdots,K$.
\item {\em Step 2 (on the physical apparatus):}  Repeat the physical process in {\em Step 
1} a number of times, $L(\epsilon,p)$, to build up a histogram of measurement frequencies 
(for all the states obtained by
measurement) until we get a probability distribution $P(T;\epsilon)$ at
the time $T$ with an accuracy $\epsilon$ for all the measured states.
The convergence of this repetition process is ensured by the Weak Law of Large Numbers in
probability theory.  (An overestimate of the number of repetitions is $L\ge 1/(\epsilon^2(1-p))$.)
Note the lowest energy state so obtained, $|\vec{n}_c\rangle$, as the candidate ground state.
\end{itemize}

Any normalised (or normalisable) state in general has a finite
support, that is, in its expansion in any complete basis of the infinite
space the magnitudes of the coefficients tend to zero as their indices
tend to infinity.  We are interested in the ground state
but in general we cannot mathematically (on (Turing) classical computer) 
determine the size of the error due to truncation of the basis. 
But we have the physical world available out there helping us to estimate the 
size of the truncation error by direct comparison with the measured
probability distributions of the last {\em Steps}.  
Once the truncation size
can be thus obtained, we can control and estimate, with some probability (confidence) $p$,
the fluctuations of the numerically obtained ground state to confirm whether 
the ground state energy is zero or not.
Explicitly,
\begin{itemize}
\item  {\em Step 3 (on the classical computer):}  Choose a
truncated basis of $M$ vectors made up of $|\alpha_1\cdots\alpha_K\rangle$ and its
excited states by successive applications of the displaced creation
operators $b^\dagger_i \equiv (a^\dagger_i -\alpha^*_i)$ on the initial state. 
\item {\em Step 4 (on the classical computer):}  Solve the 
Schr\"odinger equation in this basis for $\psi(T)$, with the initial state
$\psi(0) = |\alpha_1\cdots\alpha_K\rangle$, to derive a probability
distribution $P_{\rm est}(T;M)$ (through $|\langle\psi(T)|\cdots n_i
\cdots\rangle|^2$)
which is similar to that of {\em Step 2} and which
depends on the total number $M$ of vectors in the truncated basis.
\item {\em Step 5 (on the classical computer):}  If the two probability distributions are not uniformly
within the desired accuracy, that is, $|P_{\rm est}(T;M) -
P(T;\epsilon)|>\epsilon$, we enlarge the truncated basis by increasing the size $M$ 
and go back to the {\em Step 4} above.
\item  {\em Step 6 (on the classical computer):}  If the two probability distributions are uniformly
within the desired accuracy, that is, $|P_{\rm est}(T;M) -
P(T;\epsilon)|<\epsilon$, then use this truncated basis to diagonalise $H_P$ to yield, within an 
accuracy which can be determined from $\epsilon$, the approximated ground state $|g'\rangle$ and its
energy $E_{g'}$.
\item {\em Step 7 (on the classical computer):} 
We can now estimate in this truncated basis the gap between the groundstate and the first excited
state.  From this gap, we can make use of the quantum adiabatic 
theorem and choose a time $T$ such that 
the probability to have the system mostly in the ground state
\[ \left||\langle g'|\psi(T)\rangle|^2-1\right|<\epsilon. \]
We then go back to {\em Step 1} with this choice of $T$, which is
to amplify and thus confirm the candidate ground state as the real
ground state.
\end{itemize}

$E_{g'}$, the energy of the
numerically obtained ground state, is approximately the
same as the $H_p$-energy $E_c$ associated with
the candidate ground state $|\vec{n_c}\rangle$ obtained in measurement,
\[ E_{g'} = E_c + \delta\]
Note that $E_c$ can be assumed to be non-zero; otherwise, if it is zero then we will have found a 
solution for the Diophantine equation and can stop the algorithm then.
The fluctuation $\delta$ in  
$E_{g'}$ due to the truncation can also be numerically estimated as a function of $\epsilon$ -- it can be
easily shown that the fluctuation $\delta(\epsilon)$ tends to zero as $\epsilon$ goes to
zero.  Now, to confirm that $E_{g'}$ is in fact non-zero we only need to
enforce the condition
\begin{eqnarray}
|\delta(\epsilon)| &<& E_c,
\end{eqnarray}
by choosing $\epsilon$ in a self-consistent manner.  From this
we can conclude (with probability $p$) that $E_{g'}>0$ and thus the Diophantine equation has no root.  

To illustrate that the statistics, influenced by the spectral flow,
are sufficient for the identification purpose, 
let us assume that the lowest-energy state $|\vec{n_c}\rangle$ is the one
obtained physically $t=T$ but is not the true ground state.
After choosing a truncated basis, the computation on a classical computer
for some evolution time, which is a parameter of both the numerical
computation and the physical process, and will have two distinct scenarios:
\begin{enumerate}
\item Either the numerical computation clearly shows that $|\vec{n_c}\rangle$ 
is not even the ground state in this truncated basis.
That is, we numerically find another state $|g'\rangle$ which has lower energy (wrt $H_P$)
than that of $|\vec{n_c}\rangle$.
\item Or the numerical seems to be consistent that this state $|\vec{n_c}\rangle$
is the ``ground state" in our arbitrarily truncated basis.
\end{enumerate}
In the first case, we can immediately eliminate $|\vec{n_c}\rangle$
for it is not even the ``ground state" in a truncated space, let alone in 
the infinite space.  In the second case, we repeat both the numerical computation 
and the physical process but with
longer evolution times, chosen in {\em Step 7} above.  The numerically so-computed probability for
obtaining the state $|g'\rangle$, even though initially agrees with (within some accuracy), 
will have to eventually depart from the physically obtained
probability.  This is a consequence of the quantum adiabatic theorem.  
In the case of our classically simulated but truncated world, the 
state $|g'\rangle$ {\em is} the ground state and thus the obtaining probability will, 
as asserted by the theorem, increase with the times evolved (for evolution 
times greater than some scale).  But if the 
state $|\vec{n_c}\rangle$ is {\em not} the true ground state in the infinite world as we 
have assumed on the other hand, its measured probability 
must decrease with the times evolved (as the probability for the true 
ground state would increase for evolution times greater 
than some scale).

We only discuss here the changing rate of probabilities as functions 
of evolved time, but note that this rate is
only part of the much richer body of information obtainable from
the statistics.  The full body of information should and could, if needed, be exploited 
further for the identification of the true ground state. 

The proof of convergence and of veracity this algorithm is to be
presented elsewhere.  

In summary, to solve the Hilbert's tenth problem we need {\em both} the physical 
adiabatic processes to obtain a candidate state {\em and} the numerical Quantum
Mechanics to verify this is the ground state through the usual
statistical predictions from the Schr\"odinger equation with a few low-lying energy 
states of ${\cal H}(t/T)$.  This way, we can overcome the 
problem of where to truncate the infinite basis for a numerical study of Quantum Mechanics,
and reconcile with the Cantor diagonal arguments which state that 
the problem could not be solved entirely in the framework of classical computation. 

The key factor in the ground state verification is {\em the probability distributions}, 
which are measurable in practice, as mentioned in Sec~\ref{subsec:measure}.
However, in using the probability distributions as the verifying criteria, we have 
to assume that Quantum Mechanics is able to describe Nature correctly to the precision 
required.  Note also that we have here a peculiar situation in which the computational 
complexity, that is, the computational time, might not be known exactly {\em before}
carrying out the quantum computation -- although it can be estimated approximately.

%The most general form of partial recursive functions --> arithmetic
%propositions --> Hilbert's tenth problem.

%Quantum algorithms for the tenth problem.

\subsection{Simple illustrations {(\em in collaboration with Andrew Rawlinson)}}
In recapitulation, we need to be able to obtain and identify
the ground state of some quantum Hamiltonian which is of infinite dimensions.  
On the one hand, we cannot implement the infinity of dimensions of the 
Hilbert space in the framework of Turing computation.  On the other
hand, in the physical, finite-time process governed by the time-dependent Hamiltonian 
${\cal H}(t/T)$ we will always obtain some state at $t=T$; but how to
identify the end states?  We have argued that we need the measurability of the probability distributions.
These distributions are the end products of the {\em dynamical process}, that is, of
the time evolution through which the measured values are obtained.
Quantum mechanically, the probability distributions can be deduced from
the amplitudes which are the inner products of the state of the
system at that time (the time of measurement) with various eigenstates of
the observable (the Hamiltonian operator $H_P$ in our situation here).

Let us take the example of the simple linear equation
\begin{eqnarray}
x - 6 &=& 0.
\label{linear}
\end{eqnarray}
The spectral flows of the corresponding ${\cal H}(t/T)$ are given in
Figures 1a, 1b and 1c for various truncations of the infinite
dimensions.  The spectral flows 
in the Figures are clearly different.  
This difference implies different measurement probabilities and statistics 
when the ground state of $H_I$ is the starting state for the evolution process. 
On the classical computer, one would have to be content with a truncated
version of the infinite-dimensional Fock space.  The whole issue of
mathematical (non)computability is where to truncate! 
A comparison between the computed statistics (computable to any desired accuracy
with ever expanding truncated bases) and the measured statistics 
(measurable to any desired accuracy by repeating the experiments a necessary 
number of times) will enable us to single out the real ground state of
$H_P$.  

In this extremely simple example, Figures 1b and 1c show the ground state 
$|n_x\rangle = |6\rangle$ with early truncation of the basis.
Further addition of higher number states to the basis
does not, as expected, alter the general characteristics of the
spectral flow of these low-lying states but only helps sharpen the numerical 
values for the various statistics. 

Let us take another example of the Pythagoras equation
\begin{eqnarray}
(x+1)^2 + (y+1)^2 - (z+1)^2 &=& 0,
\label{pythagorus}
\end{eqnarray}
of which the corresponding spectral flows are presented in Figures 2a, 2b and 2c for
various truncations of the infinite dimensions.  The Figure 2c shows
the two solutions $(2,3, 4)$ and $(3,2,4)$ of~(\ref{pythagorus})
in a sufficiently large truncated basis.  In general, the information 
where to sufficiently truncate the basis will have to be provided 
by the physical process with appropriately found $T$, not by mathematical arguments alone.
Note that
we do not discuss here the complicated issue of degeneracy in the ground state of $H_P$. 
% (when the Diophantine equation has multiple solutions).  

\section{Discussion of finer points}\label{sec:dis}
\subsection{Difference from classical algorithms}
We do not look for the zeroes of the polynomial,
$D(x_1,\cdots,x_K)$, which may not exist, but instead search for the absolute 
minimum of its square which always exists,
\begin{eqnarray}
0\le \min \left(D(x_1,\cdots,x_K)\right)^2 \le \left(D(0,\cdots,0)\right)^2,
\end{eqnarray}
and is finite because $\lim_{x\to\infty} \left(D(x_1,\cdots,x_K)\right)^2$
diverges.

While it is equally hard to
find either the zeroes or the absolute minimum in classical computation, 
we have converted the problem to the realisation of the ground state of
a quantum Hamiltonian and there is no known quantum principle against
such an act.  In fact, there is no known physical principles 
against it.  Let us consider the three laws of thermodynamics concerning 
energy conservation, entropy of closed systems and the unattainability
of absolute zero temperature.  The energy involved in our algorithm
is finite, being the ground state energy of some Hamiltonian.  The
entropy increase which ultimately connects to decoherence effects is a
technical problem for all quantum computation in general.

It may appear that even the quantum process can only explore a
finite domain in a finite time and is thus no better than a classical
machine in terms of computability.  But there is a crucial
difference:
\begin{itemize}
\item  In a classical search, even if the global minimum is encountered, it cannot 
generally be 
proved that it is the global minimum (unless it is a zero of the Diophantine 
equation).  Armed only with mathematical logic, we would still have to 
compare it with all other numbers from the 
infinite domain yet to come, but we obviously can never complete this 
comparison in finite time -- thus, mathematical noncomputability.
\item  In the quantum case, the global minimum is encoded in the ground
state.  Then, by energetic tagging, the global minimum can be found in finite 
time and confirmed, if it is the ground state that is obtained at the 
end of the computation.  And the ground state may be identified and/or verified by 
physical principles.  These principles are over and above 
the mathematics which govern the logic of a classical machine and help
differentiate the quantum from the classical.  Quantum mechanics could 
``explore" an infinite domain, but only in the sense that it can select, 
among an infinite number of states, one single state (or a subspace in case 
of degeneracy) to be identified as the ground state of some given Hamiltonian
(which is bounded from below).  
This ``sorting" can be done because of physical 
principles which are not available to mathematical computability.  Note
that infinite dimensions are common in Quantum Mechanics, for example it
is well known that no finite-dimensional matrices can satisfy the
commutator
\[ [x,p] = i\hbar.\]
\end{itemize}

\subsection{Difference from the standard model of Quantum Computation} 
Our proposal is in
contrast to the claim in~\cite{QTM} that quantum Turing machines
compute exactly the same class of functions, albeit 
perhaps more efficiently, which can be computed by classical Turing machines.  
We can only offer here some speculations about
this apparent discrepancy.  The quantum Turing machine approach is a
direct generalisation of that of the classical Turing machines but with qubits
and some universal set of one-qubit and two-qubit unitary gates to build up,
step by step, dimensionally larger, but still dimensionally finite unitary operations.  
This universal set is chosen on its
ability to evaluate any desirable classical logic function.
Our approach, on the other hand, is from the start 
based on infinite-dimension Hamiltonians acting on some Fock space
and also based on the special properties and unique status of their ground states.  
The unitary 
operations are then followed as the Schr\"odinger time evolutions.  Even at the 
Hamiltonian level higher orders of the operators $a$ and $a^\dagger$, i.e. 
not just two-body but many-body interactions in a sense, are already present.  
This proliferation, which is even more pronounced at the level of the time-evolution 
operators, together with the infinite dimensionality and the unique 
energetic status of the vacuum could be the reasons behind the 
ability to compute, in a finite number of steps, what the dimensionally
finite unitary operators of 
the standard quantum Turing computation cannot do in a finite number of steps. 
Note that it was the general Hamiltonian computation that was discussed 
by Benioff and Feynman~\cite{benioff, feynman} in the conception days of 
quantum computation.

Indeed, Nielsen~\cite{nielsen} has also found no logical contradiction
in applying the most general quantum mechanical principles to the
computation of the classical noncomputable, unless certain Hermitean
operators cannot somehow be realised as observables or certain unitary
processes cannot somehow be admitted as quantum dynamics.  And up to now
we do not have any evidence nor any principles that prohibit these kinds
of observables and dynamics.

\subsection{Possible pitfalls}
% quantum optics as an example
% energy: finite, photons
% Peter Drummond's
Our quantum algorithm is based on the key ingredients of:
\begin{itemize}
\item  The exactitude, to the level required, of the theory of Quantum
Mechanics in describing and predicting physical processes;
\item  Our ability to physically implement certain Hamiltonians having infinite
numbers of energy levels;
\item  Our ability to physically obtain and verify some appropriate
ground state.
\end{itemize}
If any of these is not achievable or approximable with controllable
accuracy, the quantum algorithm simply fails and further modifications may or may not work.

For example, we could implement the algorithm with Quantum Optical
apparatuses, in which a beam of quantum light is the physical system on which final
measurements are performed and the number of photons is the quantity measured.  
The Hamiltonians could then be physically
simulated by various components of mirrors, beam splitters, 
Kerr-nonlinear media (with appropriate efficiency), etc ...  We
should differentiate the relative concepts of energy involved in this case; 
a final beam state having one single photon, say, could correspond to a 
{\em higher} energy eigenstate of $H_P$ than that of a state having more 
photons!  Only in the final act of measuring photon numbers, the 
more-photon state would transfer more energy in the measuring device 
than the less-photon state.

A fundamental problem~\cite{drummond} is that the 
Hamiltonians which we need to be simulated in the optical apparatuses are only 
{\em effective Hamiltonians} in that their descriptions are only valid 
for certain range of number of photons.  When there are too many photons, 
a mirror, for example, may respond in a different way from when only a 
few photons impinge on it, or the mirror simply melts.  That is, other 
more fundamental processes/Hamiltonians different than the desirable 
effective Hamiltonians would take over beyond certain limit in photon numbers.  
Thus it seems that we cannot have available an unbounded number of 
levels for the quantum algorithm.

This situation is not unlike that of the required unboundedness of the
Turing tape.  In practice, we can only have a finite Turing
tape/memory/register; and when the register is overflowed we would need
to extend it.

Similar to this, we would have to be content with a
finite range of applicability for our simulated Hamiltonians.  But we should
also know the limitation of this applicability range and be able to tell when
in a quantum computation an overflow has occurred -- that is, when the
range of validity is breached.  We could
then use new materials with extended range of (photon number) applicability or
modify the Diophantine equations until the appropriate ground state is
verifiably obtained.  A simple way to reduce the number of photons
involved would be a shifting the unknowns
in the Diophantine equations by
some integer amounts, $x_i \to x_i + n_i$, as in the example of~(\ref{fermat}).

This way of patching results in some approximation for our
algorithm.  The important thing is that the approximation 
is {\em controllable} because, within some given 
accuracy, only a finite number of low-lying energy levels of $H_P$
can influence the statistics for obtaining the ground state .  
In other words, the probability distribution, necessary 
for the verification of the ground state, is still measurable to within
any given accuracy with appropriate increase in the physically available
number of low-lying levels.

Controllability of the environmental effects is also another crucial 
requirement for the implementation of our quantum
algorithm.  The coupling with the environment causes some 
fluctuations in the energy levels.  Ideally, the size of these
fluctuations should be controlled and reduced to a degree that is
smaller than the size of the smallest energy gap such as not to
cause transitions out of the adiabatic process.  This may present a
difficulty~\cite{nielsenprivate} for our algorithm, but one difficulty of 
technical nature rather than of principle.  Even though these
fluctuations in the energy levels are ever present and cannot be reduced
to zero, there is no physical principle, and hence no physical reason,
why their size cannot be reduced to a size smaller than some required
scale.  The situation is similar, and closely related, to the fact that
even though we cannot attain the temperature of absolute zero, there is 
nothing in principle preventing us from approaching as closely as desired to
the absolute zero. 

However, we may not need to physically reduce the fluctuations in 
the energy levels to that degree.  They may only need to be
understood and incorporated well into our numerical estimation for 
a realistic estimate of the transition probabilities.

Only physical experiments actually performed, nevertheless, could definitely 
and finally settle the implementation issues discussed in this Section or elsewhere.

\section{Implications for other decision problems and G\"odel's Incompleteness 
Theorem}\label{sec:qcothers}
% Applications of Diophantine equations: reducibility, Tiling, Post's problem,
% Goldbach' conjecture, Riemann Hypothesis
% Incompleteness:  Proof outside the system --> truth.  (If verification is
% possible then it is a closed case.)  Can quantum proof be Godel numbered?? 
% Quantum result as oracles?
Generalised noncomputability and undecidability set the boundary for computation 
carried out by
mechanical (including quantum mechanical) processes, and in 
doing so help us to understand much better what can be so computed.
With this in mind, we mention here the considerations~\cite{kieu2} 
about some modified version of 
the Hilbert's tenth problem and about the computation of Chaitin's $\Omega$ number 
of Sec.~\ref{sec:others}.  
These problems are all inter-related through 
questions about existence of solutions of Diophantine equations.

As pointed out in Sec.~\ref{sec:others}, we can also ask different 
questions~\cite{davis} whether some Diophantine equation has a finite number 
of non-negative integer solutions (including the case in which it has no solution) or 
an infinite number, or whether the number of solutions is even, etc ...  
In general, we cannot tell the degree of degeneracy of the 
ground state; but with some modifications, the quantum algorithm above for the
Hilbert's tenth problem can be generalised to tackle this new class of
questions.  The possibility of such a generalisation confirms the
mathematical equivalence between the Hilbert's tenth problem and this new class.

Notwithstanding this, the situation is different with the evaluation of Chaitin's
$\Omega$ number.  We now have to appeal to the hypothetical ability to physically construct
Hamiltonians involving a countably infinite number of
distinct pairs of creation and annihilation operators.  (Recall that for
the Hilbert's tenth problem and its equivalence, we only need as many
pairs as the number of unknowns in a Diophantine equation.)  In other
words, we would have to possess the ability to create or simulate
Hamiltonians in Quantum Field Theory.  That is, if we stick to the
example of quantum computation with Quantum Optics, we would have 
to have individual control over infinitely many different optical modes, each 
with a distinct frequency.  Clearly, the situation is worse than before
and is even more hypothetical.

The difficulty we encounter for the $\Omega$ number is nothing but 
another manifestation of the different classes in the hierarchy of
classical noncomputability.  In Sec.~\ref{sec:others} we pointed out that, 
on the one hand, the evaluation of this number is as classically 
noncomputable as the completeness of Arithmetic and that, on the other hand, 
G\"odel's Incompleteness is ``more" noncomputable than the Hilbert's tenth 
problem.

Our decidability study so far only deals with the property of being Diophantine,
which does not cover the property of being arithmetic in general (which could involve
unbounded universal quantifiers).  As such, our consideration 
has no direct consequences on G\"odel's Incompleteness theorem.
However, it is conceivable that G\"odel's theorem may lose its
restrictive power once the concept of mathematical proof is suitably generalised with
quantum principles.

\section{Back to Mathematics}\label{sec:reform}
While the proposal
is about some quantum processes to be implemented physically, it 
illustrates the surprisingly important r\^ole of Physics in the 
study of computability.  This is an unusual state
of affairs when Physics, which has its roots in the physical world out there,
could perhaps help in setting the limits of Mathematics.

Inspired by this connection between the two, we present next
a reformulation of the Hilbert's tenth problem.  
The reformulation is made possible since physical theories in general, and
Quantum Mechanics in particular, have enjoyed the support and rigour of
mathematical languages.  We wish to stress here that, in spite of the
inspiration, the connection is established entirely in the domain
of Mathematics; we need not appeal to some real physical processes as we do 
with the proposed quantum algorithms above.

Mathematically, all we need to do is to sort out the instantaneous ground 
state $|g\rangle$ among the infinitely 
many eigenvectors of ${\cal H}(S)$ in~(\ref{Hamiltonian}); but this is a hard
task.  The trick we will use~\cite{kieu3}, as inspired by quantum adiabatic processes,
is to tag the state $|g\rangle$ by some other known state $|g_I\rangle$ 
which is the ground state of some other operator $H_I$ and can be
smoothly connected to $|g\rangle$ through some continuous parameter 
$s\in[0,1]$.  To that end, we consider the interpolating operator~(\ref{Hamiltonian}), 
rewritten as 
\begin{eqnarray}
{\cal H}(s) &=& H_I + f(s)(H_P - H_I),\nonumber\\
&\equiv& H_I + f(s)W,
\label{1}
\end{eqnarray}
which has an eigenproblem at each instant $s$, 
\begin{eqnarray}
[{\cal H}(s) - E_q(s)]|E_q(s)\rangle = 0, &&q = 0, 1, \cdots
\label{eigen}
\label{2}
\end{eqnarray}
with the subscript ordering of the sizes of the eigenvalues, $|E_0(1)\rangle \equiv |g\rangle$,
and $f(s)$ 
not necessarily linear but some continuous and monotonically increasing function in $[0,1]$
\begin{eqnarray}
f(0) = 0; && f(1) = 1.
\label{f}
\end{eqnarray}
Clearly, $E_0(0) = E_I$ and $E_0(1) = E_g$.
It turns out that for the linkage $E_0(s)$ to connect a ground state to 
another ground state we require that
\begin{eqnarray}
[H_P, H_I] &\not=& 0.
\label{symm}
\end{eqnarray}

The details are presented in~\cite{kieu3} to arrive at
\begin{eqnarray}
\frac{d}{ds}|E_q\rangle &=& f'(s)\sum_{l\not = q}^\infty \frac{\langle
E_l|W|E_q\rangle}{E_q -E_l} |E_l\rangle,
\label{10}\\
\frac{d}{ds} E_q(s) &=& f'(s)\langle E_q(s)| W | E_q(s)\rangle.
\label{4}
\end{eqnarray}
Equations~(\ref{10}) and~(\ref{4}) form the set of infinitely coupled differential 
equations providing the tagging linkage as desired.

Analytical and numerical methods
can now be employed to investigate the unknown ground state of $H_P$ from the 
constructively known spectrum of $H_I$ as the initial conditions.  In this reformulation, 
the Diophantine equation in consideration has at least one integer solution if and only if
\begin{eqnarray}
\lim_{s\to 1} E_0(s) &=& 0.
\label{answer}
\end{eqnarray}
The limiting process is necessary here since $H_P$, i.e. ${\cal H}(1)$, will have a degenerate spectrum
because of certain symmetry ($H_P$ commutes with $a^\dagger_ia_i$).
%Formally,
%\begin{eqnarray}
%E_q(s) &=& E_q(0) + \langle E_q(0)|W|E_q(0)\rangle s
%+\nonumber\\
%&& +2\sum_{l\not = q}^\infty {\rm Re}\left\{\int_0^s dt \int_0^t d\tau \frac{\langle
%E_l(\tau)|W|E_q(\tau)\rangle\langle E_q(0)|W|E_l(\tau)\rangle}{E_q(\tau) -E_l(\tau)}
%\right\}+
%\label{e0}\\
%&& +\sum_{l,m\not = q}^\infty \int_0^s dt \int_0^t d\tau \int_0^t d\tau'\frac{\langle
%E_q(\tau)|W|E_m(\tau)\rangle\langle E_l(\tau')|W|E_q(\tau')\rangle
%\langle E_m(\tau)|W|E_l(\tau')\rangle}{\left(E_q(\tau) -E_m(\tau)\right)\left(E_q(\tau') -E_l(\tau')
%\right)}.\nonumber
%\end{eqnarray}

The equations above are infinitely coupled and cannot be solved
explicitly in general.  But we are only interested in certain
information about the ground state.  And since the influence on the ground state 
by states having larger and larger indices diminishes more and more thanks to
the denominators in~(\ref{10}) (once no degeneracy is assured), 
this information may be derived, numerically or 
otherwise, with some truncation to a finite number of states involved. 
This may not work for all the Diophantine equations. 

While the ground-state outcome for our differential equations might or
might not be computable, it should be noted that there are 
instances~\cite{barrow,ill-posed} 
where very simple differential equations, such as the wave equations,
could have noncomputable outcomes because of ill-posed initial conditions.

With care we can slightly modify the derivation
for~(\ref{10},\ref{4}) to come up with similar equations even when there is some
degeneracy in $[0,1]$.  But for the condition~(\ref{answer}) to be the
indicator for the existence of solutions of the Diophantine equation, simple
topological consideration only requires
{\it that the initial ground state $|E_0(0)\rangle= |g_I\rangle$ is not degenerate and 
that this state does not cross with any other state in the open interval $s\in(0,1)$.}  
With the freedom of choice for $H_I$  satisfying~(\ref{symm}), we should be able to 
eliminate any symmetry in the open interval $s\in(0,1)$ for ${\cal H}(s)$ in order to 
have a stronger condition of a totally avoided crossing.

\section{Concluding remarks}
%Implications:...
%Halting Problem as a special case of the Godel Incompleteness Theorem,
%now the halting is quantum decidable what then is the status of Godel's
%theorem in the quantum logic??
We have discussed and emphasised the 
issue of computability in principle, not that of computational complexity.  
This attempt of broadening the concept of effective computability, taking
into account the quantum mechanical principles, has been argued to be able in
principle
to decide some of the classical undecidables, the Hilbert's 
tenth problem and thus the Turing halting problem in this instance. 
If the quantum algorithm is realisable, and we do not have any evidence of fundamental nature
to the contrary, the Church-Turing thesis for effective computability
should be modified accordingly.  But first, in need of
further investigations are the effects of errors in the implementation of
Hamiltonians, and of decoherence, and of measurement on the final outcome of 
our algorithm.

On the other hand, if for any reasons the algorithm is not implementable 
{\it in principle} then it would be an example of information being limited by
physics, rather than by logical arguments alone.
That is, there might be some
fundamental physical principles, not those of practicality, which prohibit
the implementation.  Or, there might not be enough 
physical resources (ultimately limited by the total energy and the lifetime 
of the universe) to satisfy the execution of the quantum algorithm.  (In this case 
it is likely that the Turing program under consideration, even if it eventually 
halts upon some input, would take a running time longer than the lifetime 
of the universe.)  In either case, the whole exercise is still very 
interesting as the unsolvability of those problems and the limit of mathematics
itself are also dictated by physical principles and resources.

That some generalisation of the notion of computation could help solve 
the previous undecidability/noncomputability has been recognised before %~\cite{peter}
in mathematics.  % and was considered by Kleene as quoted in~\cite{recursive}. 
But quantum physics has not been recognised as the missing ingredient until very recently.  
Our quantum algorithm presented here
could in fact be regarded as an infinite search through the integers in a 
finite amount of time, the type of search required to solve the Turing halting problem.  
Apart from ours, there also exist in the literature some other efforts where physical
principles are explored~\cite{nz, relativity, nielsen, stannett} for some possible
extension of the notion of computability.

%%%
% Stress on measurability and computability.
It should be emphasised again here that not only the values of some
observables are measured but so are the probability distributions of
these values.  These measurable probability distributions are also
to be compared against those obtained from the theory of Quantum Mechanics. 
This helps identify the ground state, and thus the answer for the decision problem, 
and is the pivotal element of our algorithm to separate quantum computability from Turing
computability.  In doing so, we have to assume that Nature is describable by Quantum
Mechanics correctly at least to the precision required.  If this is not
the case, testing a mathematically solvable
Diophantine equation might yield some evidence for the failure of Quantum 
Mechanics as a theory of Nature.

Also implied in the discusions above is the fact that Quantum Mechanics {\em is},
as a theory, non-computable.  This fact is not so widely recognised.

To understand quantum computability and its limits, we have also
considered, with mixed results, some problems generalised from the Hilbert's 
tenth and the evaluation of Chaitin's $\Omega$ number.  Nevertheless,
our study is an illustration of ``Information is physical".  

Inspired by Quantum Mechanics, we have also reformulated the question of
solution existence of a Diophantine equation into the question of certain
properties contained in an infinitely coupled set of differential equations.  
In words, we encode the answer of the former question into the smallest
eigenvalue and corresponding eigenvector of a hermitean operator bounded from below.
And to find these eigen-properties we next deform the
operator continuously to another operator whose spectrum is known.
Once the deformation is also expressible in the form of a set of coupled
differential equations, we could now start from the constructive
knowns as a handle to study the desired unknowns.

Note that the reformulation is entirely based on mathematics. 
If a general mathematical method could be found to extract the required
information from the differential equations for any given Diophantine equation
then one would have the solution to the Hilbert's tenth problem itself.  
This may be unlikely but not be as contradictory as it seems --because the 
unsolvability of the Hilbert's tenth problem is only established in the 
framework of integer arithmetic and in Turing computability, not necessarily 
in Mathematics in  general.  Tarski~\cite{tarski}  has shown that the question 
about the existence of {\it real} solutions of polynomials over the reals is, 
in fact, {\it decidable}.

It seems appropriate to end here with a quotation from the man whose famous
Incompleteness result has often been misquoted as spelling the end for 
computability/provability in Arithmetic.  In G\"odel's own words~\cite{godelw}: 
\begin{quotation}
``... On the other hand, on the basis of 
what has been proved so
far, it remains possible that there may exist (and even be empirically
discoverable) a theorem-proving machine which in fact {\em is} equivalent
to mathematical intuition, but cannot be {\em proved} to be so, nor even
be proved to yield only {\em correct} theorems of finitary number theory.''
\end{quotation}
Perhaps, quantum computation {\it is} that possibility?  (In the affirmative case, 
the notion of proof may also need to be reconsidered,
see~\cite{kieurandom} and references therein.)

\section*{Acknowledgements}  I am indebted to Alan Head and Peter Hannaford
for discussions, comments and suggestions.  I would also like to thank Cristian 
Calude, Jack Copeland, Bryan Dalton, John Markham, Michael Nielsen,
Boris Pavlov, Andrew Rawlinson, Falk Scharnberg and Khai Vu for discussions; 
Gregory Chaitin, Peter Drummond, Gabor Etesi, Yuri Matiyasevich, Ray Sawyer, Karl Svozil,
Boris Tsirelson and Ray Volkas for email correspondence.  Comments by anonymous referees are 
also gratefully acknowledged. 
I wish to acknowledge a Visiting Scientist position and
the hospitality extended to me during my stay at the CTP at MIT, and
also a visit at the IAS at Princeton.  Discussions with Stephen Adler, Enrico Deotto, Edward Farhi,
Jeffrey Goldstone, and Sam Gutmann have helped me sharpening up the
arguments presented herein.

% \newpage
\begin{quotation}
{\em Tien D Kieu} came to Australia as a Vietnamese refugee in 1980 and
obtained a B Sc (Hons) from the University of Queensland in 1984 and
a Ph D from the University of Edinburgh in 1988.
He then held postdoctoral positions at the Universities of Edinburgh and
Oxford, including a Junior Research Fellowship at Linacre College.  In
1991, he took up an Australian Postdoctoral
Fellowship and then a Queen Elizabeth II Fellowship at the University of Melbourne.  
At the beginning of 2001, after three years at CSIRO in the position of 
Senior Research Scientist and
Project Leader, he joined the Swinburne University of Technology as a Professorial 
Fellow.  He is currently interested in Quantum Computation and Quantum Measurement.
\end{quotation}
%\newpage
\noindent
{\bf Figures 1:}  The spectral flow of ${\cal H}(s)$, with $\alpha=0.5$ and
with $s\equiv t/T$ is the scaled evolution time, corresponding to the linear 
equation~(\ref{linear}) in the truncated bases with 5, 9, and 25 states respectively in
Figures 1a, 1b and 1c. Figures 1b and 1c contain the unique solution $|n_x\rangle=|6\rangle$.
Enlargement of the truncated basis from here on
would not change the characteristics of the spectral flow of these low-
lying states,
but only approximates better the measurement probabilities as functions of
evolution time.

\noindent
{\bf Figures 2:}  The spectral flow of ${\cal H}(s)$, with $\alpha=0.5$ and with $s\equiv t/T$ 
is the scaled evolution time, 
corresponding to the Pythagoras 
equation~(\ref{pythagorus}) in the truncated bases with 84, 220, 1771 states respectively in 
Figures 2a, 2b, and 2c.  Figure 2c contains two
solutions of the Pythagoras equation~(\ref{pythagorus}),
$|n_x, n_y, n_z\rangle = |2, 3, 4\rangle$ and $|3, 2, 4\rangle$.  Further
enlargement of the truncated basis will increase the degeneracy of the
ground state.

\end{document}